\documentclass[a4paper, oneside, twocolumn, notitlepage, 10pt]{extarticle_ecoc}
\usepackage{ecoc}

\usepackage[nolist]{acronym}
\usepackage{orcidlink}
\begin{document}
\selectlanguage{english}    


\title{DRL-Assisted Dynamic QoT-Aware Service Provisioning in Multi-Band Elastic Optical Networks}%


\author{
    Yiran Teng\orcidlink{0009-0003-3740-691X}\textsuperscript{(1)}, 
    Carlos Natalino\orcidlink{0000-0001-7501-5547}\textsuperscript{(2)},
    Farhad Arpanaei\orcidlink{0000-0003-1061-0614}\textsuperscript{(3)},\\
    Alfonso S\'anchez-Maci\'an\textsuperscript{(3)}, 
    Paolo Monti\orcidlink{0000-0002-5636-9910}\textsuperscript{(2)},
    Shuangyi Yan\orcidlink{0000-0002-5021-2840}\textsuperscript{(1)},
    Dimitra Simeonidou\textsuperscript{(1)}
}

\maketitle                  




\begin{strip}
    \begin{author_descr}

        \textsuperscript{(1)} High Performance Networks Group, Smart Internet Lab, University of Bristol, Bristol, United Kingdom,
        \textcolor{blue}{\uline{shuangyi.yan@bristol.ac.uk}}

        \textsuperscript{(2)} Dept. of Electrical Engineering, Chalmers University of Technology, 412 96 Gothenburg, Sweden,

        \textsuperscript{(3)} Dept. of Telematic Engineering, Universidad Carlos III de Madrid, 28911 Leganes, Madrid, Spain.

    \end{author_descr}
\end{strip}

\renewcommand\footnotemark{}
\renewcommand\footnoterule{}


\begin{strip}
    \begin{ecoc_abstract}
        We propose a DRL-assisted approach for service provisioning in multi-band elastic optical networks. Our simulation environment uses an accurate QoT estimator based on the GN/EGN model. Results show that the proposed approach reduces request blocking by 50\% compared with heuristics from the literature.
        \textcopyright 2024 The Author(s)
    \end{ecoc_abstract}
\end{strip}


\section{Introduction}


Quality of transmission (QoT)-aware dynamic service provisioning in multi-band elastic optical networks (MB-EONs) comprises routing, modulation format selection, band allocation, and spectrum assignment (RMBSA). The RMBSA problem is usually divided into several sub-problems, which are addressed primarily through heuristics and/or machine learning (ML)-based methods.
Rule-based heuristics \cite{Sambo:2020:jlt, Calderón:2022:icl, Jana:2022:jocn, Yao:2022:jlt} are based on fixed human-designed rules (e.g., shortest path routing), which are not primarily designed for multi-band systems and lack flexibility.
Among the ML-assisted solutions, reinforcement learning (RL) has been reported as a viable technique for reducing blocking probability (BP) by using the Q-table-based method for service provisioning \cite{Terki:22:ecoc, Terki:23:psc}. 
However, Q-table suffers from scalability issues in complex and/or continuous observation-action spaces such as those experienced in RMBSA and lacks generalization abilities.
To this end, deep RL (DRL)-assisted RMBSA algorithms have been investigated \cite{Sheikh:2021:ondm, Gonzalez:2022:latincom, Beghelli:2023:ondm}. 
DRL employs deep neural networks (DNNs) to extract essential features from complex environments, which handle continuous observation spaces and generalize well. 
Although DRL has shown great potential in service provisioning within single-band optical networks \cite{Chen:2019:jlt, tang2022heuristic, Etezadi:2023:jocn, Teng:2023:psc}, its effectiveness in the MB scenario has not been demonstrated, i.e., DRL did not outperform simple heuristics, e.g., $k$-shortest-path first-band first-fit (KSP-FB-FF) \cite{Sheikh:2021:ondm, Gonzalez:2022:latincom, Beghelli:2023:ondm}.
This is mainly because these studies merely extended the DRL-based model designed for single-band scenarios \cite{Chen:2019:jlt} without considering some critical MB features such as fragmentation and band utilization. 
For instance, the adoption of fragmentation metrics has shown benefits in single-band EONs \cite{tang2022heuristic, Etezadi:2023:jocn}.
Additionally, as the number of bands increases, achieving a high-performance DRL model becomes more challenging \cite{Beghelli:2023:ondm}. For instance, the expansion of the action space significantly affects the capability of the DRL agent to learn effective policies.

In this paper, we propose a novel DRL-assisted QoT-aware RMBSA algorithm for MB-EONs. 
The DRL agent encompasses several innovations.
First, the MB-EON environment incorporates a QoT estimator based on the Gaussian noise (GN)/enhanced GN (EGN) model to estimate the channel-connection profile for all route/band/channel alternatives. Consequently, the bit rate/modulation format for each lightpath is contingent upon the corresponding channel QoT value. This is in contrast to prior works such as \cite{Sheikh:2021:ondm, Gonzalez:2022:latincom, Beghelli:2023:ondm} which adopted a (less accurate) distance-adaptive approach. 
Second, the design of the DRL agent leverages a novel state representation tailored for MB-EONs, which enhances the agent's ability to capture essential network information. 
Third, action masking filters invalid actions, thereby reducing the DRL agent's decision space. 
Simulation results demonstrate that the proposed DRL-based solution reduced the BP by around 50\% compared to the heuristics from the literature.
To the best of our knowledge, this DRL-assisted QoT-aware RMBSA for dynamic MB-EONs based on GN/EGN model is the first of its kind.
\section{Physical Layer Model}
\label{sec:pli}

This paper focuses on L+C+S-band EONs, where an aggregate bandwidth of 20 THz (6+6+8) is divided into 268 channels, each with 75 GHz (6 $\times$ 12.5 GHz), considering a 400 GHz gap between adjacent bands.   
Since the bandwidth exceeds 15 THz, the approximated triangle model for the Raman gain profile is not applicable to the considered system model \cite{Semrau2022}. Therefore, we apply a GN/EGN semi-closed form model to estimate the non-linear interference (NLI) noise, including the Kerr effects and the inter-channel stimulated Raman scattering (ISRS) \cite{RanjbarECOC22022}. The model takes into account frequency-dependent physical layer parameters such as loss, dispersion, and non-linear coefficients. Therefore, the actual Raman gain profile, dependent on the pump frequency and the frequency offset, is obtained by solving the coupled Raman differential equations. In the model, the modulation format and dispersion correction terms are adopted to improve its accuracy. The validation of this model was conducted through field trials \cite{Yanchao_ECOC_2023}. In contrast to previous related works such as \cite{Terki:22:ecoc, Terki:23:psc, Sheikh:2021:ondm, Gonzalez:2022:latincom, Beghelli:2023:ondm}, we utilize the power optimization method introduced in the literature \cite{FARHAD_OFC_2024}, which accounts for amplified spontaneous emission (ASE) noise loading in idle channels \cite{Arpanaei2023ONDM}. According to the incoherent GN model for uncompensated optical transmission links \cite{Poggiolini2014}, the GSNR for a lightpath (LP) on channel $i$ can be derived from \eqref{eq:GSNR_total} as follows:
\begin{multline}\label{eq:GSNR_total}
	GSNR^{i}_{\text{LP}}|_{\text{dB}} = 10 \log_{10} \Big[\big(\sigma_{\text{ASE}}+\sigma_{\text{NLI}}+
 \sigma_{\text{TRx}}^{-1}\big)^{-1} \Big]\\-\sigma_{\text{Flt}}|_{\text{dB}} - \sigma_{\text{Ag}}|_{\text{dB}},
\end{multline}
where $\sigma_{\text{ASE}}= \Sigma_{s\in S}P^{s,i}_{\text{ASE}}/P_{\text{tx}}^{s+1,i}$ and $\sigma_{\text{NLI}}= \Sigma_{s\in S}P^{s,i}_{\text{NLI}}/P_{\text{tx}}^{s+1,i}$. Moreover, $P_{\text{tx}}^{s+1,i}$ is the launch power at the beginning of span $s+1$, $P^{s,i}_{\text{ASE}}= n_{\text{F}}hf^i(G^{s,i}-1)R_{\text{ch}}$ is noise power caused by the doped fiber amplifier (DFA) equipped with the digital gain equalizer, and NLI noise power ($P^{s,i}_{\text{NLI}}$) is calculated from equation (2) in the reference \cite{RanjbarECOC22022}. Moreover, 
$n_{\text{F}},\,h,\,f^i,\,G^{s,i}=P_{\text{tx}}^{s+1,i}/P_{\text{rx}}^{s,i},\, S$, and $R_{\text{ch}}$ are the DFA's noise figure, Plank’s coefficient, channel frequency, frequency center of the spectrum, DFA's gain, set of spans, and channel symbol rate, respectively. $P_{\text{rx}}^{s,i}$ is the received power at the end of span $s$. $\sigma_{\text{TRx}}$, $\sigma_{\text{Flt}}$, $\sigma_{\text{Ag}}$ are the transceiver SNR, SNR penalty due to wavelength selective switches filtering, and SNR margin due to the aging. Moreover, the Xponders in the MB-EON are equipped with state-of-the-art flexible bit rate/modulation format transceivers (TRxs). The modulation format of TRxs can adaptively change based on the QoT.

\section{DRL for QoT-Aware RMBSA}
We design a DRL agent for service provisioning in MB-EONs.
Upon the arrival of request $R_{t}(s,d,b) $ at time step $t$, the DRL agent must select a route between node $s$ and $d$, choose a band, and assign a number of channels for $R_t$ that satisfies the requested bit rate $b$ using the first-fit scheme.
The objective of the DRL agent is to minimize the long-term BP under the QoT/GSNR constraint.
Below, we elaborate on the design of the DRL agent, including the specific configurations of the state ($s_t$), action ($a_t$), and the reward ($r_t$).
\noindent \textbf{State space ($s_t$):} 
In our case, the $s_t$ is a $K \times (|E| + 5\times B)$ vector, where $K$, $|E|$, and $B$ are the number of candidate routes, links, and bands, respectively.
For each route among $K$ candidates, $|E|$ elements $\in \{0,1\}$ are utilized to indicate which links are used on this route (1 for used links).
Next, for each band on this route, the candidate channel(s) (determined by the first-fit scheme) in the band with sufficient capacity to accommodate the bit rate requested by $R_t$ are obtained.
Four features are extracted from candidate channel(s) in each band:
\textbf{\textit{(i)}} number of the channel(s) necessary to support the requested bit rate,
\textbf{\textit{(ii)}} maximum bit rate supported by the channel(s) when using their highest modulation format,
\textbf{\textit{(iii)}} average index of the channel(s),
\textbf{\textit{(iv)}} number of free channels on the adjacent links that occupy the same spectrum as these channel(s) (related to fragmentation).
The last feature, i.e., \textbf{\textit{(v)}}, is the total bit rate supported by the available channels of each band on this route (related to band utilization).  
\noindent \textbf{Action space ($a_t$):} The $a_t$ is a $K \times B + 1$ vector corresponding to the probability of selecting each action. These actions encompass choosing one of the $K \times B$ candidate path/band or rejecting the $R_t$.
An action masking scheme \cite{Huang:2020:arXiv} is employed to prevent the selection of invalid actions (i.e., candidates without sufficient channels to accommodate $R_t$).
This approach has shown that the agent can enhance performance \cite{shimoda2021mask} and focus on more important features \cite{Ayoub:2024:xrl}.
The mask sets the probabilities of selecting invalid actions to 0.
\noindent \textbf{Reward ($r_t$):} The $r_t$ is 1 if the request $R_t$ is provisioned, -1 otherwise.
\begin{figure}[t]
    \centering
    \includegraphics[width=0.99\linewidth]{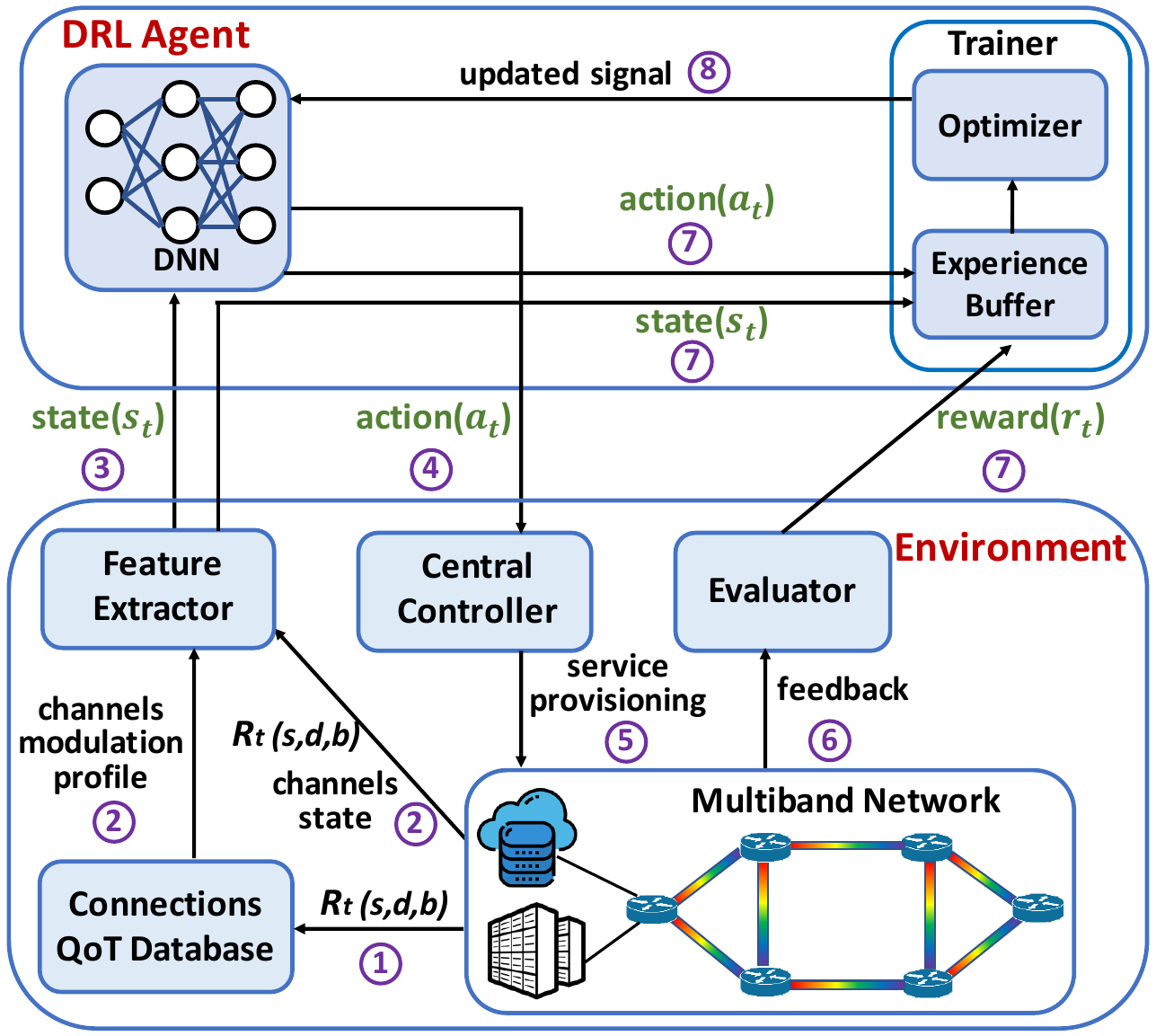}
    \captionsetup{justification=centering}
    \caption{Framework and workflow of the DRL-assisted QoT-aware service provisioning solution in multi-band optical networks.}
    \label{Fig1}
\end{figure}
The framework and the workflow of the proposed solution are depicted in Fig. \ref{Fig1}.
Upon receiving a request $R_t$, the channels' modulation profile of the KSPs is obtained by traversing the pre-computed Connections QoT Database (step 1). 
This profile, the request information, and the channels' state are sent to the Feature Extractor (step 2), which generates the $s_t$ (step 3). 
In turn, $s_t$ is fed to the DNN of the DRL agent which returns the $a_t$ (step 4).
Next, the Central Controller provisions the service based on $a_t$ (step 5).
This provisioning result is sent to the evaluator which calculates $r_t$ (step 6).
Subsequently, the tuple ($s_t$, $a_t$, $r_t$) is stored in the Experience Buffer as a training sample (step 7).
Once the Experience Buffer is full, the Optimizer triggers a training epoch, updating the DNN parameters (biases and weights) according to the gradient descent method (step 8).
Through iterative cycles of these processes, i.e., episodes, the DRL agent progressively refines its provisioning policy to maximize the discounted reward, which, in our case, minimizes the long-term BP.

\section{Results}
The simulations refer to the NSFNET topology and using L+C+S-band. 
Six modulation formats from PM-BPSK to PM-64QAM at 64 Gbaud are considered.
The connections QoT database adopts the GSNR thresholds for each modulation format defined in the literature\cite{6DMANJOCN2024}, corresponding to a pre-forward error correction bit error rate of 1.5e-2. Therefore, each channel has a bandwidth of 75 GHz and can provide bit rates ranging from 100 to 600 Gb/s when using PM-BPSK to PM-64QAM, respectively.
We consider Erbium-DFAs (EDFAs) with noise figures of 4.5 dB and 5 dB for the C- and L-band, respectively, and a Thulium-DFA (TDFA) with a noise figure of 6 dB for the S-band. We assume a standard single mode fiber with a zero-water peak.
The spectrum continuity is considered for channel assignment along a route.

The MB environment was established based on the Optical RL-Gym\cite{2020:Natalino:ICTON}.
The advantage actor-critic (A2C) \cite{Mnih:2016:ICML} algorithm was used for training. 
The hyperparameters of the DRL agent are set as follows. The discounted factor $\gamma$, learning rate, training buffer size, and mini-batch size are 0.95, 5e-5, 1,000, and 500, respectively. The size of the DNN is $5 \times 128$ (5 layers and 128 neurons), and $ReLU$ is chosen as the activation function.  
The proposed solution was compared with heuristics including $K$-shortest-path first-band first-fit (KSP-FB-FF) \cite{Sambo:2020:jlt}, distance-adaptive first-fit (KSP-DA-FF) \cite{Calderón:2022:icl}, and bit-rate-adaptive first-fit (KSP-BA-FF)\cite{Calderón:2022:icl}.
We do not include previously proposed DRL solutions \cite{Sheikh:2021:ondm, Gonzalez:2022:latincom, Beghelli:2023:ondm} because they cannot outperform existing heuristic \cite{Sambo:2020:jlt}.

\begin{figure}[!t]
    \centering
    \includegraphics[width=0.96\linewidth]{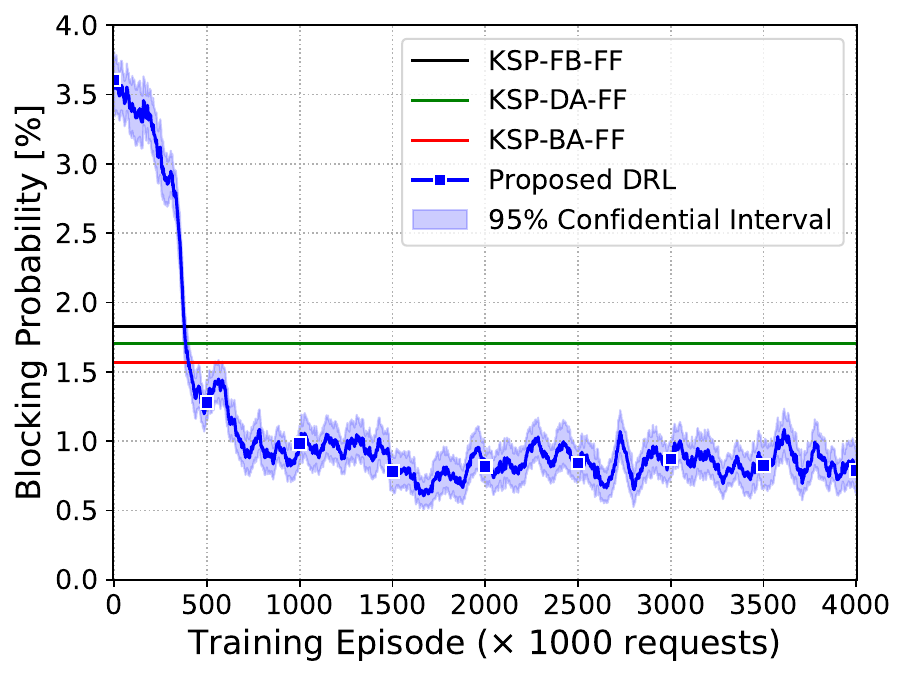}
    \captionsetup{justification=centering}
    \caption{BP over training episodes of different solutions in L+C+S muti-band network under 800 Erlang}
    \vspace{3mm}
    \label{Fig2}
\end{figure}

\begin{figure}[!t]
    \centering
    \includegraphics[width=0.96\linewidth]{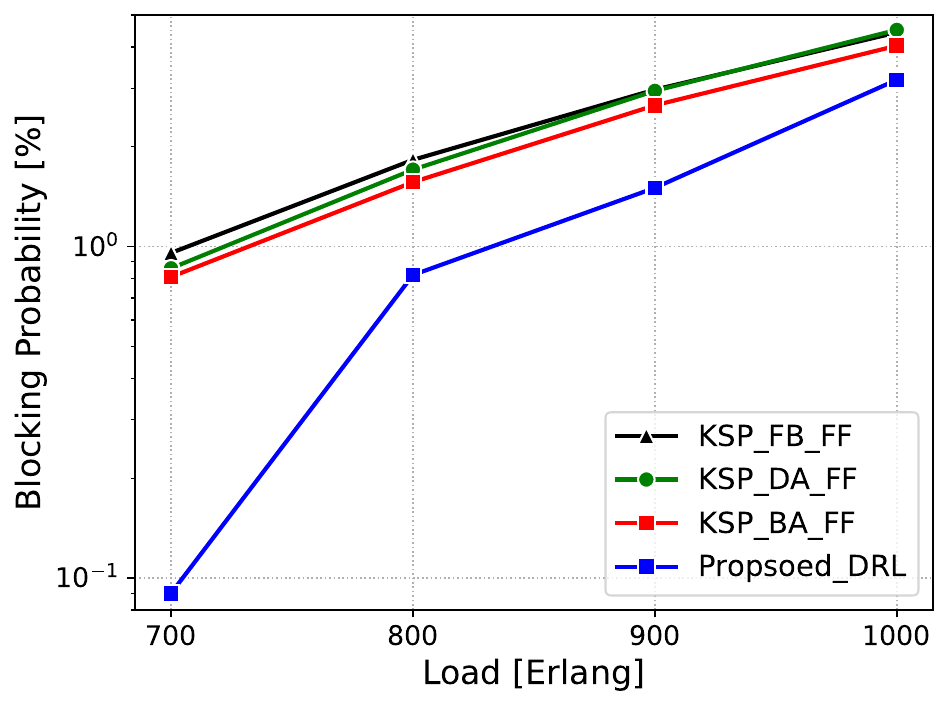}
    \captionsetup{justification=centering}
    \caption{BP of different solutions in L+C+S muti-band network under different traffic loads}
    \label{Fig3}
\end{figure}

Figure \ref{Fig2} shows the BP (smoothed over a 75-episode window) achieved by the DRL agent during training.
At the initial stage of the training, the DRL agent is initialized with random parameters and achieves a poor BP.
However, as the training progresses, the DRL agent continually optimizes its policy, which is reflected in a significant decrease in BP.
After around 1,500 training episodes, the performance of the DRL agent reaches a steady performance.
At this stage, the DRL agent can reduce the BP by 55\%, 51\%, and 47\% compared to KSP-FB-FF, KSP-DA-FF, and KSP-BA-FF, respectively.
The performance evaluation across various traffic loads, shown in Fig. \ref{Fig3}, validates the effectiveness of the proposed solution.
The benefits of the DRL-based solution become more pronounced under lower traffic loads.
Specifically, the BP reduction achieved by the DRL agent is around 90\% under 700 Erlang. 
Compared to previous DRL-based solutions \cite{Sheikh:2021:ondm, Gonzalez:2022:latincom, Beghelli:2023:ondm} in MB-EONs, our approach shows two key advantages. 
Firstly, our state representation is carefully designed for the MB environment, containing features that are highly related to the BP of the network.
This state representation enables the agent to capture crucial information about the network and effectively explore the environment.
Secondly, incorporating action masking also contributes to the improvement. 
Action masking eliminates the need for the agent to learn how to differentiate unavailable actions, simplifying the DRL agent exploration process and leading to more effective use of the features\cite{Ayoub:2024:xrl}.

\section{Conclusions}
In this paper, we proposed a DRL-assisted QoT-aware RMBSA algorithm for dynamic service provisioning in MB-EONs.
A GN/EGN QoT estimator was employed to create the route-channel modulation format profiles for the DRL agent.
This solution outperforms several presented heuristics in reducing BP in the L+C+S-band scenario.

\clearpage
\section{Acknowledgements}
The authors acknowledge the support of EU-funded projects -- Allegro (No. 101092766) and ECO-eNET (No. 10113933.) -- and the UK EPSRC project HASC (No. EP/X040569/1). F. Arpanaei acknowledges support from the CONEX-PLUS project under the Marie Sklodowska-Curie Action program (grant No. 801538).

\defbibnote{myprenote}{%
Citations must be easy and quick to find. More precisely:
\begin{itemize}
    \item Please list all the authors. 
    \item The title must be given in full length. 
    \item Journal and conference names should not be abbreviated but rather given in full length.
    \item The DOI number should be added incl. a link.
\end{itemize}
}
\printbibliography

\vspace{-4mm}

\acrodef{ASE}{amplifier spontaneous emissions}
\acrodef{CUT}{channel under test}
\acrodef{DRL}{deep reinforcement learning}
\acrodef{EGN}{enhanced Gaussian noise}
\acrodef{GSNR}{Generalized Signal-to-Noise Ratio}
\acrodef{NLI}{non-linear impairments}

\end{document}